# NMR Study in the Iron-selenides $Rb_{0.74}Fe_{1.6}Se_2$: Determination of the Superconducting Phase as Iron Vacancy-Free $Rb_{0.3}Fe_2Se_2$

Y. Texier,[1] J. Deisenhofer,[2] V. Tsurkan,[2,3], A. Loidl,[2] D. S. Inosov,[4] G. Friemel,[4] J. Bobroff[1,*]

[1]*Laboratoire de Physique des Solides, Univ. Paris-Sud, UMR8502, CNRS, F-91405 Orsay Cedex, France*
[2]*Experimental Physics V, Center for Electronic Correlations and Magnetism, University of Augsburg, 86135 Augsburg, Germany*
[3]*Institute of Applied Physics, Academy of Sciences of Moldova, MD-2028 Chişinău, Republic of Moldova*
[4]*Max Planck Institute for Solid State Research, Heisenbergstraße 1, D-70569 Stuttgart, Germany*

$^{77}$Se and $^{87}$Rb nuclear magnetic resonance (NMR) experiments on $Rb_{0.74}Fe_{1.6}Se_2$ reveal clearly distinct spectra originating from a majority antiferromagnetic (AF) and a minority metallic-superconducting (SC) phase. The very narrow NMR line of the SC phase evidences the absence of Fe vacancies and any trace of AF order. The Rb content of the SC phase is deduced from intensity measurements identifying $Rb_{0.3(1)}Fe_2Se_2$ as the actual composition of the SC fraction. The resulting estimate of 0.15 electrons/Fe brings this class of superconductors 245 family closer to the other Fe-based superconductor families.

Iron-based superconductors are often described in the limit of moderate to weak electronic correlations [1]. The parent compounds are usually antiferromagnetic (AF) semimetals with small moments, multiorbital band structure and display good nesting between hole and electron pockets, suggesting an itinerant spin-density wave scenario. When the nesting is weakened, a superconducting (SC) state develops, as observed under electron or hole doping for example. But the recent discovery of superconductivity in the 245 iron-selenide family has put this scenario into question. Indeed, this family of compounds with nominal composition $A_xFe_{2-y}Se_2$ (with $A$ = K, Rb, Cs) shows antiferromagnetism with large moments of 3.3 $\mu_B$, ordering temperatures $T_N$ up to 550 K, and an insulating behavior accompanied by Fe vacancy order. Unexpectedly, this AF order seems to coexist with superconductivity with $T_c$ of about 30 K [2]. Furthermore, the Fermi surface does not seem to have any hole pocket, which prohibits a nesting scenario [3]. These findings would favor a strong coupling scenario instead, where the physics could be captured by the carrier-doping of a Mott insulator as in high $T_c$ cuprates [4]. This may also imply a different superconducting pairing symmetry than in other Fe-based superconductors [5].

However, the experimental situation is much more complex and many recent reports have argued that superconductivity and antiferromagnetism are in fact spatially separated [6-11]. In recent scanning tunneling microscopy (STM) on thin films, phase separation between AF and SC regions was observed in $K_xFe_{2-y}Se_2$ [6]. An ARPES photoemission study also measured the electronic signature in the reciprocal space of an insulating phase mesoscopically separated from a superconducting one [7]. But STM and ARPES techniques probe only the surface of the material where specific reconstruction effects could occur, especially in the presence of Fe vacancies. Local probes (μSR, Mössbauer, or NMR) sensitive to the magnetism in the bulk volume of the material are especially well suited to detect any phase separation. μSR [8] and Mössbauer [9] studies found that the 245 Rb or K materials segregate into spatially separated SC and AF regions, as also suggested by TEM experiments [13]. But surprisingly, $^{87}$Rb and $^{77}$Se [14, 15] NMR studies both reported the observation of just one single NMR line which displays all the characteristics of a pure metallic-SC phase, and did not observe any experimental evidence for phase separation. Two main questions arise from this puzzling body of evidence: Why is NMR at odds with other local probes? And even more crucial, what is the nature of the metallic-SC phase? Does it feature the same characteristics as other Fe-based superconductors? To answer this, experiments are needed to measure its stoichiometry and doping content, and if it contains Fe vacancies similar to the AF phase. This issue is essential to settle the debate about the role and strength of the correlations in Fe-based superconductors.

In this Letter, we report $^{87}$Rb and $^{77}$Se NMR experiments providing clear evidence for the phase separation and we explain why previous NMR studies missed it. We further demonstrate by analyzing the spectral shapes of the lines of both nuclei that the metallic-SC phase is surprisingly homogeneous and does not contain any Fe vacancies nor magnetic moments. $^{87}$Rb NMR allows us to directly determine the Rb stoichiometry of this phase, which corresponds to 0.15 electrons/Fe doping. We conclude that this SC phase is similar to other Fe-based superconductors and that 245 selenides should not be described in a Mott scenario.

We studied a superconducting $Rb_{0.74}Fe_{1.6}Se_2$ single crystal. Details about synthesis and characterization can be found in [16], where the sample used in the present experiment is labeled “BR26”. Macroscopic measurements show that it is both AF below $T_N \approx 600$K and SC below $T_c \approx 32$ K. The sample was always kept in helium or nitrogen atmosphere to avoid degradation. NMR was performed in a field-sweep experiment around $H$ = 14 T (for $^{77}$Se) and 7.5 T (for $^{87}$Rb). All measurements reported were done with applied magnetic fields parallel to the crystallographic $c$ axis. We also performed measurements with the field perpendicular to the $c$ axis which display similar qualitative features. All the spectra were obtained by standard Fourier transform recombination. Relaxation $T_1$ measurements were performed in a saturation-recovery sequence. The superconducting transition was observed in-situ by the detuning of the NMR coil at $T_c$ = 27 K at 7 T and at 25 K at 14 T.

Typical $^{87}$Rb NMR spectra are plotted in Fig. 1. They consist of a set of three narrow lines and a broad background, which corresponds to the SC and AF phases, respectively, as proven hereafter. The $^{77}$Se NMR displays only a narrow component due to the SC phase and no broad signal related to the AF phase, as shown in Fig. 2(a).

Let us first focus on these narrow components plotted in Fig. 2(a) which were already observed in previous reports [12,14,15]. As $^{87}$Rb is a spin 3/2 nucleus, the spectrum consists of a central line (nuclear transition 1/2↔-1/2) and two quadrupolar satellites (nuclear transitions ±3/2↔±1/2) due to the local electric field gradient (EFG). Since the $^{77}$Se nucleus carries a spin ½, it is not sensitive to the EFG and displays just a central transition. Figure 2(c) shows the temperature dependence of the shift of the $^{87}$Rb and $^{77}$Se central lines versus temperature. Both shifts display the same behavior, which is proportional to the Fe layer spin susceptibility [16]. This proves that both nuclei are hyperfine coupled to the Fe layers. The corresponding spin dynamics probed by the longitudinal relaxation time $T_1$ shows a Korringa-like behavior typical of the normal metallic state in these materials. The $^{87}$Rb and $^{77}$Se Knight shifts [Fig. 2(c)] and $1/T_1$ [Fig. 2(d) red points] are found to drop at $T_c$ on decreasing temperatures. This demonstrates that these narrow lines are related to Rb and Se sites in a SC phase. This is confirmed by the sharp increase of the linewidths below $T_c$ [(Fig. 2(b)] which signals the development of a field distribution due to the appearance of a vortex state, as analyzed in details in [14].

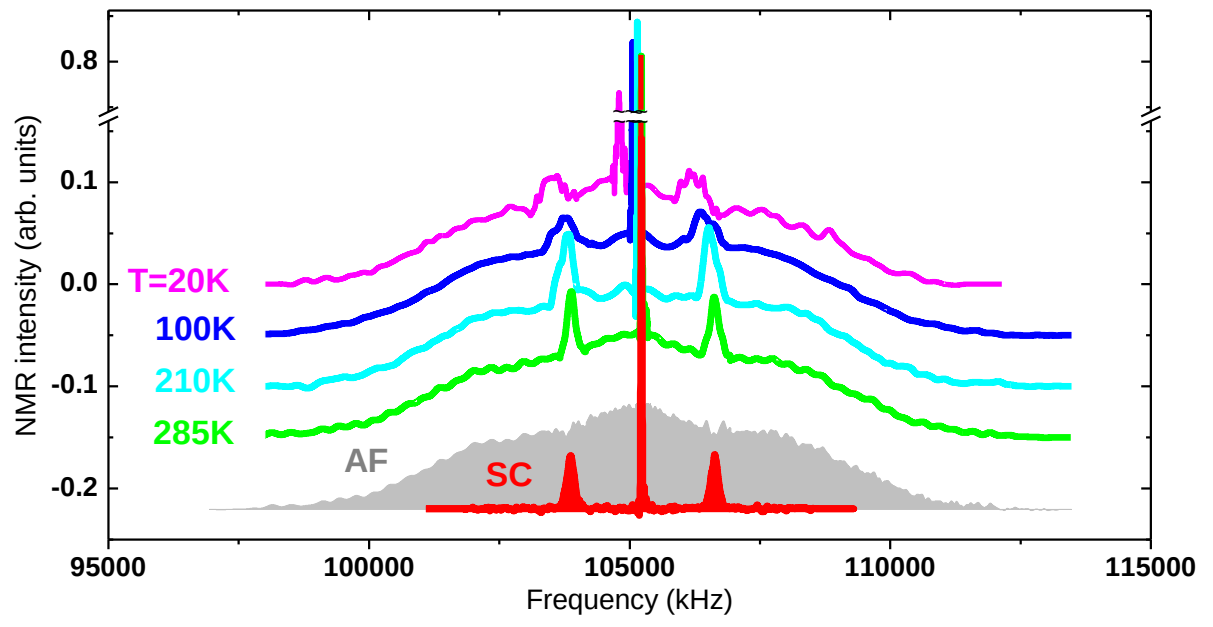

Figure 1: (color) $^{87}$Rb NMR spectra for H//c at various temperatures display a narrow-superconducting (SC) and broad-antiferromagnetic (AF) contribution which can be separated by contrast measurements as shown at the bottom of the figure. Curves are vertically shifted with respect to each other.

We now turn to the broad $^{87}$Rb background seen in Fig. 1 which was not detected in previous studies. This broad line is found unchanged when decreasing temperature, even below $T_c$. Its large width of about 14 MHz and the fact that the central line cannot be distinguished from the quadrupolar satellites implies a very large internal field distribution of about 1 T, on the contrary to the SC line. This signals the presence of frozen magnetic moments on the Fe sites. Hence, we can attribute this contribution to the AF phase observed with neutron scattering measurements [10, 17]. Its low-temperature relaxation rate $1/T_1$ differs from that of the SC phase, as evidenced in Fig. 2(d). We further measured that its transverse relaxation time $T_2$ at room temperature is about 12 (1) µs, while that of the SC phase is one to 2 orders of magnitude larger (~ ms). This difference in $T_2$ allowed separating the broad and narrow spectral contributions by a contrast experiment using short and long delays $\tau$ in a $\frac{\pi}{2} - \tau - \pi$ echo sequence, as shown in Fig. 1. The observed 1 T width is compatible with local moments of 3.3 $\mu_B$ and Fe vacancies √(5)×√(5) pattern, when using the typical hyperfine coupling of Rb to FeSe layers. The constant behavior in temperature is also compatible with the fact that the AF order develops at a much higher temperature $T_N$≈600K. A more refined simulation of the NMR spectrum is beyond the scope of this work because the hyperfine coupling and EFG tensors are not known especially close to the Fe vacancies.

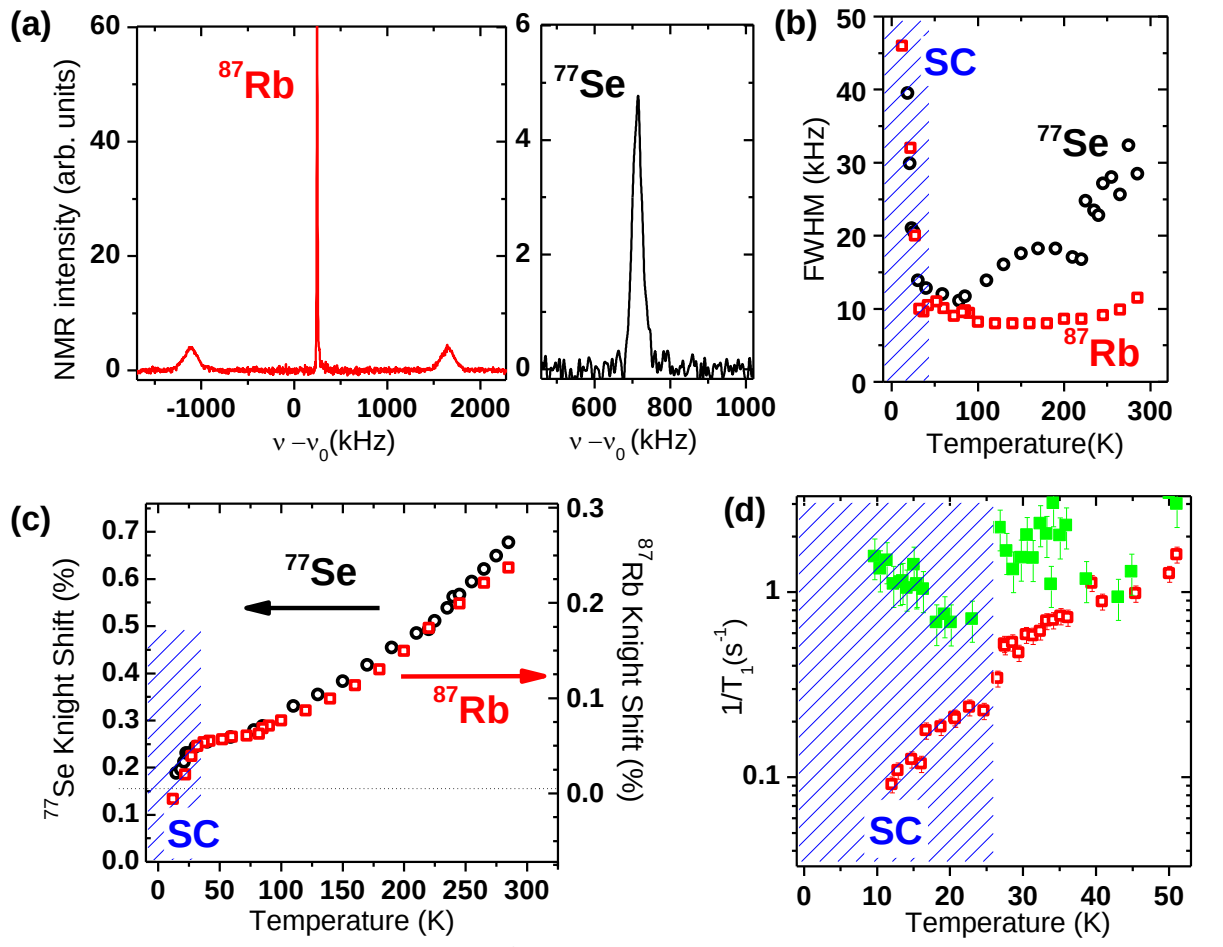

Figure 2: (color) (a) $^{77}$Se and $^{87}$Rb NMR spectra for the SC phase only (obtained by contrast experiments for Rb) for H//c. (b) Corresponding full width at half maximum. (c) Temperature dependence of the SC Knight shift of $^{87}$Rb and $^{77}$Se. (d) Temperature dependence of the relaxation rate $\frac{1}{T_1}$ for the SC (red) and AF (green) $^{87}$Rb phases. In all figures, the hatched region indicates where superconductivity occurs.

The large linewidth and the very short $T_2$ make this AF signal hard to detect, explaining why previous $^{87}$Rb NMR study missed it [12]. One might expect a similar AF broad signal in the $^{77}$Se spectrum. However, the $^{77}$Se hyperfine coupling to Fe spins is twice as large as that of $^{87}$Rb, which implies that this AF Se signal would be 2 times broader and with a transverse relaxation $T_2$ which is 4 times shorter. Such a short $T_2$ of only a few μs makes the signal impossible to detect in standard NMR pulse experiments, which explains why it has not been observed, neither by us nor by other groups [12,14,15].

Measurements of the $^{87}$Rb SC and AF spectral weights show that the SC/AF signal intensities are in proportion $4/96 \pm 0.5$. This ratio does not depend on temperature, which implies that the metallic-SC phase contains about 20-25 times less Rb atoms than the AF phase. We also compared the total $^{87}$Rb and $^{77}$Se intensities: the observed $^{77}$Se line is about 1 order of magnitude smaller than it should be according to the nominal chemical composition [18]. This confirms that the SC $^{77}$Se signal represents only a minor fraction of the total volume of the sample, the Se of the AF phase being undetected. These various results clearly confirm the μSR and Mössbauer findings of a phase separation between metallic-SC and AF regions in this material [8, 9].

Our data provide new information about these phases. In the AF phase, the low-temperature $^{87}$Rb $T_1$ (green points in Fig. 2(d) clearly differs from that of the metallic and SC phase. The data show a large scattering due to the very small signal intensity and relatively long $T_1$ values which limit the signal to noise ratio. But well beyond this experimental uncertainty, $T_1$ is about an order of magnitude smaller in the AF phase than in the metallic phase below $T_c$, strongly suggesting that this AF phase does not coexist at an atomic level with superconductivity, contrary to observations in other Fe-based superconductors [19].

On the other hand, the NMR linewidth of the metallic-SC phase above $T_c$ is surprisingly narrow as already stressed. To put it in context, it is about 4 times narrower than in the cleanest high temperature cuprate superconductors measured by NMR. This implies that the SC phase must be free of any local disorder and well separated from any frozen magnetism of the AF phase. We computed that any frozen moment on Fe site larger than 0.1μB should indeed create on the adjacent Rb nuclei a dipolar magnetic field of about 34 kHz in Rb frequency units, while the observed field distribution is only 10-20 kHz. We also computed the effect of the full AF ordered phase (assuming the $\sqrt{(5)} \times \sqrt{(5)}$ AF pattern) on the SC $^{87}$Rb line for various separation distances. The nearest neighbor and next-nearest neighbor Rb sites of the AF phase should experience a dipolar field of ~0.2 T and 0.0013 T respectively, i.e., 3000 kHz and 18 kHz in Rb frequency units. This sharp decrease with distance comes from the cancellation of the AF alternating moments. The experimental Rb linewidth is therefore compatible with a SC phase separated at least 15 Å from the AF phase (i.e., one unit cell). Recent TEM [13] and near-field microscopy [20] experiments performed on the same material suggested that this compound consists of alternating SC and AF layers [19], mimicking a natural hetero-structure arrangement as displayed in Fig. 3(a). This layered pattern is fully compatible with our data since the two phases have just to be separated by one unit cell to make the AF dipolar contribution vanish in the SC layer.

This phase separation could also be observed in a sample from the same batch by synchrotron x-ray diffraction performed at the Swiss-Norwegian beam line of the European Synchrotron Radiation Facility (ESRF). The data show splitting of both $c$-axis and in-plane Bragg reflections into two peaks, consistent with previous reports [21], whose intensity ratio agrees with the volume fractions of the two phases estimated in the present work.

The small $^{77}$Se and $^{87}$Rb linewidths also point toward the absence of any Fe vacancies in the FeSe layers in the metallic-SC phase. Indeed, any Fe vacancy should modify the NMR shift of the nearby $^{77}$Se and $^{87}$Rb nuclei because of their hyperfine coupling with the Fe atoms. Fe vacancies should lead to the appearance of a $^{77}$Se satellite line shifted by 25% less than the main line since the nearby $^{77}$Se would couple to 3 Fe and not 4.

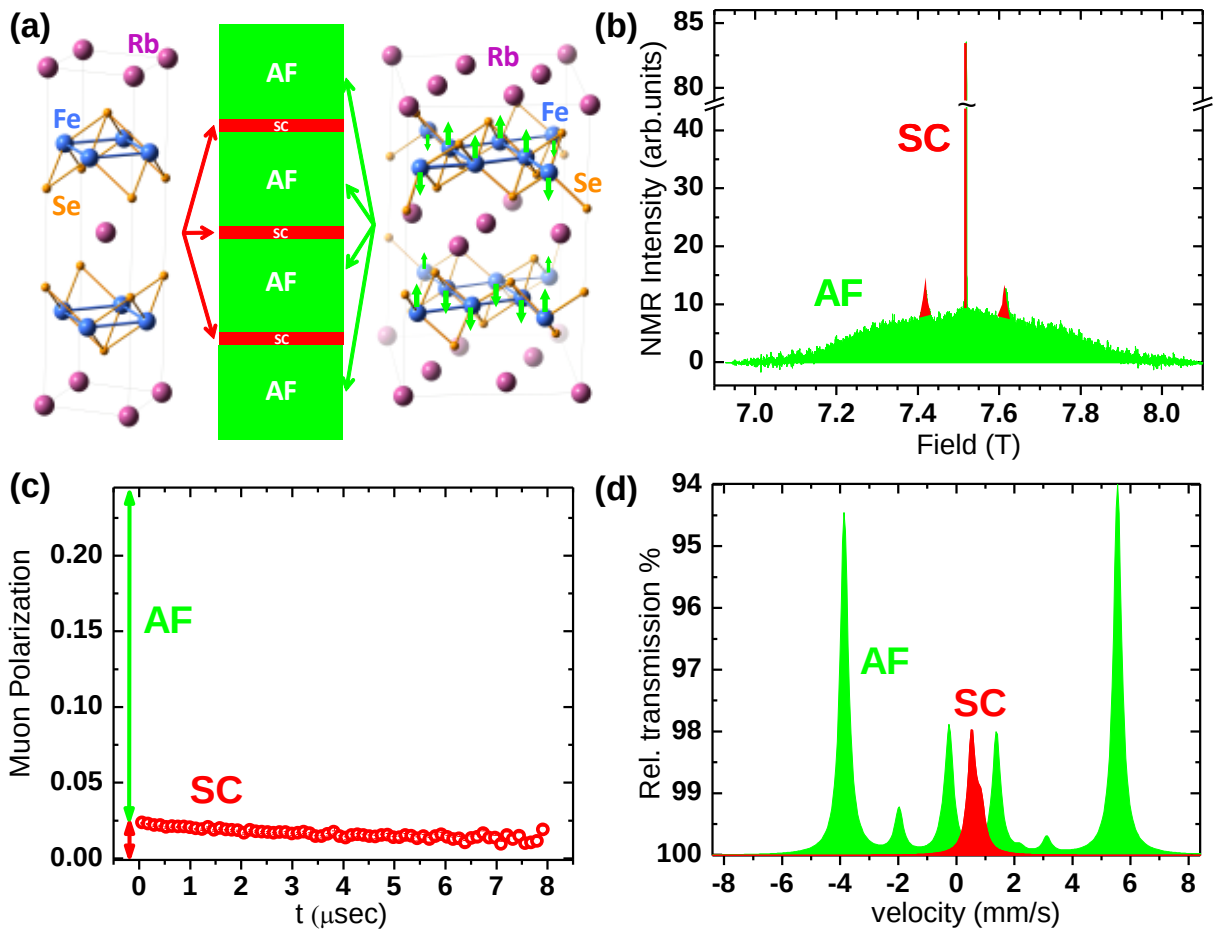

Figure 3: (color) (a) Proposition for an alternated layer arrangement of the AF $Rb_{0.74}Fe_{1.6}Se_2$ (green) and SC $Rb_{0.3}Fe_2Se_2$ (red) phases from Ref. [19]. (b), (c) and (d) the Rb NMR, μSR from Ref. [8] and Mössbauer from Ref. [9] data all show separated contributions from the SC (red) and AF (green) phases.

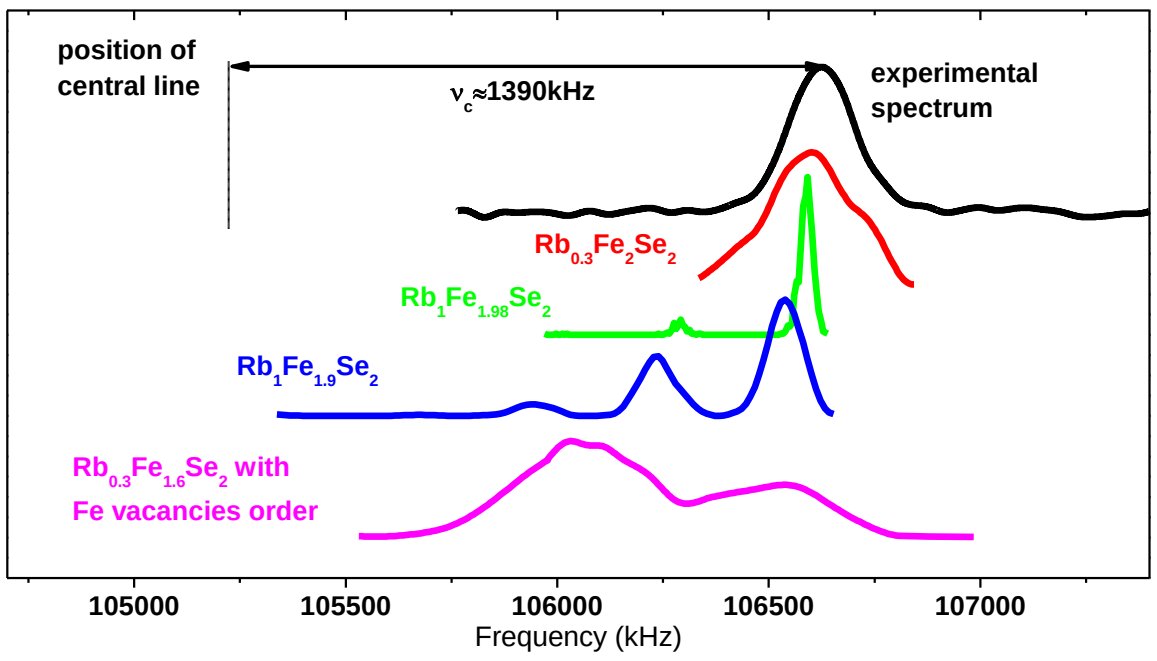

Figure 4: (color) $^{87}$Rb quadrupolar satellite (black) is compared to EFG point-charge simulations in $Rb_x^{+}Fe_{2-y}^{2+}Se_2^{2-}$ assuming randomly distributed Rb sites for x=0.3 (red) or Fe vacancies either randomly distributed (green, blue) or ordered in a **√(5)**×**√(5)** pattern (magenta).

We do not observe any indication of this satellite. Another even stronger evidence of the absence of Fe vacancies comes from the fact that the Rb quadrupolar satellites are well resolved and rather narrow. The EFG quadrupolar shift of 1.4 MHz between the central line and the satellites seen in Fig. 2(a) should be strongly modified by any Fe vacancy. We performed simulations of how the quadrupolar satellite would transform in the presence of Fe vacancies in a crude point-charge model assuming the SC phase to be $Rb_x^{+}Fe_{2-y}^{2+}Se_2^{2-}$ and rescaling the absolute values to the experimental quadrupolar shift. As plotted in Fig. 4, only a few percent of Fe vacancies split the quadrupolar satellite into many lines (green and blue spectra), in clear contradiction with our experimental observation (black spectrum). This stays true even if Fe vacancies are ordered according to the $\sqrt{(5)} \times \sqrt{(5)}$ pattern (magenta spectrum), independent of the Rb content. We conclude that the upper boundary for Fe vacancies in the metallic-SC FeSe layer must be about 1%. On the contrary, Rb vacancies (red spectrum in Fig. 4) are found not to affect too much the quadrupolar satellite, so our experiment is compatible with partially occupied Rb layers.

This spectral shape analysis lets us conclude with confidence that the FeSe layers are stoichiometric in the SC phase, i.e., with composition $Rb_xFe_2Se_2$. The Rb stoichiometry $x$ can be estimated using $^{87}$Rb spectral intensities together with the findings of μSR [8] and Mössbauer [9]. These three experiments are summarized in Fig. 3. The green/red signals represent the AF/SC phases as determined in these different experiments (see [8,9] for details). NMR, Mössbauer and μSR intensities (or spectral weights) probe respectively the number of Rb ions, the number of Fe ions and the volume fraction of the SC/AF phases. The NMR SC/AF intensity ratio of 4/96 should then be equal to the volume ratio 12/88 determined by μSR multiplied by $x$/0.74. This leads to $x$=0.29±0.06. This value is independently confirmed by the intensity of the Se NMR signal of the metallic-SC phase which should be $2/x \approx 7\pm1$ times larger than the Rb one, and is found experimentally to be ~ 8 times larger [18]. Regarding Mössbauer, our estimate implies a ratio between the number of Fe sites in the SC and AF phase of about 14/86, a value compatible with the observation made in experiments [Fig. 3(d)] within experimental uncertainties. It is also compatible with inelastic neutron scattering data [22]. This new determination of $x$ made on experimental grounds independent of any model gives new insight in the nature of the 245 family. The same reasoning allows us to determinate the AF phase Rb stoichiometry to be 0.8, a value slightly higher than the 0.74 nominal composition.

Our findings indeed show that the SC phase in the bulk of 245 iron-selenides has the composition $Rb_{0.3(1)}Fe_2Se_2$, i.e., FeSe layers with a doping of 0.15 electrons per Fe. At such a doping level, most Fe-pnictide compounds are known to exhibit superconductivity. Furthermore, the spin lattice relaxation rate $1/T_1$, the Knight shift and the linewidth in the superconducting state look very similar to other Fe-based superconductors such as $Ba(Fe_{1-x}Co_x)_2As_2$. The coexistence of superconductivity and strong local moment antiferromagnetism in the 245 iron-selenide was taken as evidence in favor of a strongly correlated Mott-insulator scenario and against a more weak-coupling approach based on an itinerant nesting picture. But we demonstrate here that the AF and SC phases segregate, and that the SC phase in the 245 compounds has nothing very specific, but is merely an electron doped iron-selenide layer without any Fe vacancies and with no local Fe moments. Thus, the Mott picture cannot be argued to explain superconductivity in the 245 iron-selenide family. These compounds remain however original and interesting as they display a natural heterostructure where superconducting layers alternate with a peculiar AF state containing ordered Fe vacancies.

We acknowledge H. Alloul, V. Brouet, P. Hirschfeld, Y. Laplace and P. Mendels for fruitful discussions, D. Chernyshov at ESRF for sample characterization, Z. Shermadani for providing μSR data, and the ANR Pnictides for support. This work has also been partially supported by the DFG within the SPP 1458, under Grant No. DE1762/1-1, and via TRR80 (Augsburg-Munich).

* julien.bobroff@u-psud.fr

[1] J. Paglione & R.L. Greene, Nature Physics **6**, 645–658 (2010).
[2] J. Guo *et al.*, Phys. Rev. B **82**, 180520 (2010).

[3] T. Qian *et al.*, Phys. Rev. Lett. **106**, 187001 (2011); Y. Zhang *et al.*, Nat. Mat. **10**, 273 (2011); D. Mou *et al.*, Phys. Rev. Lett. **106**, 107001 (2011);
[4] R. Yu, J.-X. Zhu, and Q. Si, Phys. Rev. Lett. **106**, 186401 (2011)
[5] T. A. Maier, S. Graser, P. J. Hirschfeld, and D. J. Scalapino, Phys. Rev. B **83** 100515(R) (2011); F. Wang et al., Europhys. Lett. 93, 57003 (2011); I. I. Mazin, Phys. Rev. B **84**, 024529 (2011); T. Saito, S. Onari and H. Kontani, Phys. Rev. B **83**, 140512 (2011)
[6] Li *et al*., Nat. Phys. **8**, 126 (2012)
[7] Chen *et al.*, Phys. Rev. X **1** 021020 (2011)
[8] Z. Shermadini *et al.*, Phys. Rev. B **85**, 100501(R) (2012)
[9] V. Ksenofontov *et al.*, Phys. Rev. B **84**, 180508(R) (2011)
[10] M. Wang *et al.*, Phys. Rev. B **84**, 094504 (2011)
[11] R. H. Yuan *et al.*, Scientific Reports **2**, 221 (2011) ; A. Ricci *et al.*, Phys. Rev. B **84**, 060511 (2011); A. Ricci *et al.*, Supercond. Sci. Technol. **24** 082002 (2011); A. Charnukha *et al.*, Phys. Rev. B 85, 100504(R) (2012)
[12] L. Ma *et al.*, Phys. Rev. B, **84**, 220505 (2011).
[13] Y. J. Yan et *al.*, Scientific Reports 2, 212 (2011).
[14] D. A. Torchetti *et al.*, Phys. Rev. B **83**, 104508 (2011).
[15] W. Yu *et al.*, Phys. Rev. Lett. **106**, 197001 (2011); H. Kotegawa *et al.*, J. Phys. Soc. Jpn. **80** 043708 (2011).
[16] V. Tsurkan *et al.*, Phys. Rev. B **84**, 144520 (2011).
[17] V. Yu Pomjakushin, E. V. Pomjakushina, A. Krzton-Maziopa, K. Conder and Z. Shermadini, J. Phys. Condens. Matter **23** 156003 (2011); W. Bao *et al.*, Chin. Phys. Lett. **28**, 086104 (2011).
[18] In the Se-Rb intensity comparison, various parameters change between the two nuclei which were taken into account: applied field, relaxation times, pulse conditions, EFG influence, spin value, gyromagnetic ratio and natural abundance.
[19] Y. Laplace, J. Bobroff, F. Rullier-Albenque, D. Colson and A. Forget, Phys. Rev. B **80**,140501 (2009).
[20] A. Charnukha *et al.*, arXiv: 1202.5446 (unpublished) (2011)
[21] A. Bosak *et al.*, arXiv: 1112.2569 (unpublished) (2011)
[22] J. T. Park *et al*., Phys. Rev. Lett. **107**, 177005 (2011); G. Friemel *et al.*, Phys. Rev. B **85**, 140511(R) (2012).